\documentclass{emulateapj}

\usepackage{amsmath}
\usepackage{cases}
\usepackage{txfonts}
\usepackage{tipa}
\usepackage{amsfonts}
\usepackage{amssymb}
\usepackage{graphicx}
\usepackage{multirow}
\usepackage{float}

\shortauthors{Du Cuihua et al.}

\begin{document}

\title{ The high velocity stars in the Local Stellar Halo from Gaia and LAMOST}

\author{Cuihua Du\altaffilmark{1,3}, Hefan Li\ \altaffilmark{2}, Shuai  Liu\altaffilmark{4}, Thomas Donlon \altaffilmark{3}, Heidi Jo Newberg\altaffilmark{3}}

\affil{$^{1}$College of Astronomy and Space Sciences, University of Chinese Academy of Sciences, Beijing 100049, China; ducuihua@ucas.ac.cn\\
$^{2}$School of Physical Sciences, University of Chinese Academy of Sciences, Beijing 100049,  China \\
$^{3}$Department of Physics, Applied Physics and Astronomy, Rensselaer Polytechnic Institute, Troy, NY 12180, USA, newbeh@rpi.edu\\
$^{4}$Key Laboratory of Optical Astronomy, National Astronomical Observatories, Chinese Academy of Sciences, Beijing 100012, China\\
}

\begin{abstract}
\par Based on the first Gaia data release and spectroscopy from the LAMOST Data Release 4, we study the kinematics and chemistry of the local halo stars. The halo stars are identified kinematically with a relative speed of at least 220 km s$^{-1}$ with respect to the local standard of rest. 
In total, 436 halo stars are identified.  From this halo sample, 16 high velocity (HiVel) stars are identified.  We studied the metallicity and [$\alpha$/Fe] distribution of these HiVel stars.  Though most of HiVel stars are metal-poor, there are several stars that have metallicity above $-0.5$ dex.  To understand the origin of high velocity stars, we evolve the trajectory of the star backwards along the orbit in our adopted Galaxy potential model to determine the orbital parameters and assess whether the star could have originated in the Galactic center.  We found that some high velocity stars could have originated from the Galactic center, but other stars were probably kicked up from the Galactic disk.

\end{abstract}

\keywords{Galaxy:abundance-Galaxy:halo-Galaxy:kinematics and dynamics-Galaxy:formation}

\section{Introduction}

\par High velocity (HiVel) stars, discovered in the Galactic halo \citep{Brown05, Hirsch05,Edelmann05}, 
are moving sufficiently 
fast so that they could escape from the Galaxy.  
The orbits of HiVel stars 
can provide useful information about the environments in which they are produced.  
In general, the extreme velocities of high velocity stars suggest that they were ejected from the Galactic center (GC) by the interactions of stars with a massive black hole \citep[MBH][]{Hills88} or a hypothetical binary MBH \citep{Yu03}.  For either scenario, the binary stars could be injected into the vicinity of the MBH from the young stellar disk in the GC \citep[e.g.,][]{Lu10, Zhang10} or from the Galactic bulge \citep{Perets09}.  It is also possible that high velocity stars could originate from the interaction of a BH binary with a single star \citep{Yu03}, or a star cluster \citep{Fragione16}.
Other models proposed to explain the HiVel stars that do not originate in the GC include: the surviving companion stars of type Ia supernova explosions \citep{Zubovas13, Tauris15}; the tidal debris of an accreted and disrupted dwarf galaxy \citep{Abadi09} or globular cluster; the result of the interactions between multiple stars \citep{Gvaramadze09}; and runaways ejected from the Large Magellanic Cloud \citep{Boubert16, Boubert18}.

\par  Recent studies have used the chemical and kinematic information to determine the origin of HiVel stars \citep[e.g.,][]{Wang13, Hawkins15, Li12, Geier15}. 
A few studies have used only the kinematics of HiVel stars to obtain an estimate of the Galactic mass  and Galactic escape speed\citep[e.g.,][]{Smith07, Piffl14}.
Since the first hypervelocity star was discovered by \cite{Brown05}, more than 20 hypervelocity have been found  \citep[e.g.,][]{Brown06, Brown09, Brown12, Brown14, Zheng14, Geier15, Huang17}.  Most of these are $2-4M_{\odot}$ late B-type stars in the Galactic halo.  Some studies suggest that HiVel stars are also metal-poor \citep[e.g.,][]{Schuster06}; \cite{Ryan03} studied a sample of intermediate metallicity HiVel stars and found that most of these stars resemble the stars in the thick disk.  
In order to put constraints on the origin of the HiVel stars, it is necessary to study the chemical distribution of late type HiVel stars in the local halo.   
These studies will also help to better understand the structure and formation of the Galactic halo, in which many of these HiVel stars currently reside. 
For example,  if the HiVel stars are more metal-rich ([Fe/H]$>-0.5$) than expected for the inner halo, and the [$\alpha$/Fe] measurements are consistent with those of disk stars, it may suggest  that these metal-rich HiVel stars formed in the disk and were subsequently dynamically ejected into the halo \citep{Bromley09, Purcell10}.

\par In the standard hierarchical model of galaxy formation, stellar halos are thought to form via the accumulation of stars
from infalling dwarf galaxies.  This merging process creates many stellar streams in the Galactic halo \citep{Searle78, Freeman02}.
However, there are many sources for stellar halo material besides direct accretion from infalling galaxies.  Some simulations suggest that a fraction of kinematically defined halo stars are in situ stars \citep{Zolotov09, Font11, Brook12, Cooper15} that formed in the initial collapse \citep{Samland03} of a galaxy, or `runaway' stars \citep{Boubert18} that formed in the disk and were subsequently kinematically heated \citep{Bromley09,Purcell10}.   These `runaway' disk stars are a subclass of HiVel stars that can provide important clues to Galactic halo formation.  

\par Although there is some evidence that both in situ and accreted stars are present in the Milky Way halo, 
the origin of the in situ stars is still unclear due to poorly measured proper motions and parallaxes.  However,  as ongoing and future surveys such as Gaia \citep{Perryman01} provide us with large numbers of radial velocities and proper motions of Galactic stars which are much more precise than previously available, it will be possible to construct accurate three-dimensional velocity distributions for nearly complete samples of nearby halo stars. These 3D maps allow us to identify 
the HiVel stars with higher fidelity and subsequently explore their origins. 

\par In this study, we use Gaia proper motions combined with radial velocities and metallicities derived from LAMOST stellar spectra \citep{Zhao12} to search for HiVel stars in the solar neighborhood.  
Section 2 introduces the observational data from Gaia and LAMOST, describes the sample selection, and defines the coordinate systems in the study.
In Section 3,  we kinematically split the sample into disk and halo components, and extract the local halo sample stars. 
In Section 4, we identify these rare HiVel stars in the solar neighborhood and explore their origin, including an analysis of the chemical abundances and orbital properties. The conclusions and summary are given in Section 5.

\section{Data}

Studying the kinematics and chemistry of the stellar sample requires 6D phase space information. The first year of Gaia (DR1) provides 5D measurements in the solar neighborhood; radial velocity measurements are not included. We complement the data with radial velocity and metallicity from the LAMOST survey. 

\subsection{Gaia and LAMOST}

\par  Gaia is a space-based mission which is obtaining accurate parallaxes and proper motions for more than one billion sources brighter than G $\sim20.7$.  The first the Gaia data release (Gaia DR1) was released in  September 2016 \citep{Gaia16a, Gaia16b},  and contains positions,  parallaxes and proper motions for $\sim2$ million of the brightest stars which are in the Tycho-2  catalog and have V $\sim12$ \citep{Hog00}. The joint catalog is known as Tycho-Gaia Astrometric Solution \citep[TGAS;][]{Lindegren16}.   The 5-parameter astrometric solutions for TGAS stars were obtained by combining Gaia observations with the positions and their uncertainties of the Tycho-2 stars (with an observation epoch of around J1991) as prior information. The resulting catalog has median parallax uncertainties of $\sim0.3$ mas, with an additional systematic uncertainty of about $\sim0.3$ mas \citep{Gaia16a, Lindegren16, Astra16}. TGAS parallax error is smaller than 1 mas, which is comparable to the Hipparcos precision (which has typical uncertainties of 0.3 mas in positions and parallaxes, and 1 mas/yr in proper motions). 
Most of these stars are within a few kpc from the Sun, while a few objects such as supergiants exist at distances of $\sim50$ kpc.

\par The Large Sky Area Multi-object Fiber Spectroscopic Telescope (LAMOST) is a 4 meter quasi-meridian reflective Schmidt telescope with 
4000 fibers within a field of view of $5^{\circ}$. The
LAMOST spectrograph has a resolution of R $\rm \sim$ 1,800 and
wavelength range spanning 3,700 {\AA} to 9,000 {\AA} \citep{Cui12}.
LAMOST has completed 4 years of survey operations plus a Pilot Survey, and has internally
released a total $\sim 6.08$ million spectra to the collaboration.
Of these, $\sim 4.33$ million are AFGK-type stars with
estimated stellar atmospheric parameters as well as $\rm{\alpha}$-element abundances and radial velocities. 
The survey reaches a limiting magnitude of $r=17.8$ (where $r$ denotes magnitude in the SDSS $r$-band),
but most targets are brighter than $r\sim17$.  
The scientific motivation and survey target selection are described in \citet{Zhao12}, \citet{Deng12}, and \citet{Liu14}.

\par The LAMOST Stellar Parameter Pipeline at Peking University [LSP3] \citep{Xiang15, Xiang17}
 determines atmospheric parameters 
by template matching with the MILES spectral library \citep{Sanchez06}. 
Compared to the ELODIE spectra \citep{Prugniel01} which are secured using an echelle spectrograph with a very high spectral resolution (R $\sim42 000$), the MILES spectra are obtained using a long-slit spectrograph at a spectral resolution (FWHM$\sim2.4${\AA}), which is comparable to that of the LAMOST spectra, and are accurately flux-calibrated to an accuracy of a few percent over the $3525-7410${\AA} wavelength range.   The stellar atmospheric parameters of MILES spectra have been calibrated to a uniform reference \citep{Cenarro07}. On the other hand, the radial velocities of MILES stars are not as accurately determined as those in the ELODIE library \citep{Prugniel01}, given the fairly low spectral resolution of MILES spectra. Thus for radial velocity determinations, the LSP3 continues to make use of the ELODIE library \citep{Moultaka04}.   $\alpha$-element to iron abundance ratio [$\alpha$/Fe] is a good indicator of the Galactic chemical enrichment history.  LSP3 also gives the overall  $\alpha$-element  (Mg, Si, Ca and Ti) to iron abundance ratio [$\alpha$/Fe]\citep{Li16, Xiang17}.  

For LSS-GAC spectra of FGK stars of SNRs per pixel higher than 10, the current implementation of LSP3 has achieved an accuracy of 5.0 km/s, 150 K, 0.25 dex, and 0.15 dex for the radial velocity, effective temperature, surface gravity and metallicity, respectively.  To provide a realistic error estimate for $\rm{[\alpha/Fe]}$, the random error induced by spectral noises is combined with the method error, which is assumed to have a constant value of 0.09 dex, estimated by a comparison with high-resolution measurements.  The detailed description of the parameters determination can be found in \citet{Xiang15, Xiang17}.

\subsection{Sample selection and Coordinate Systems}

\par The data used in our work are from two catalogues; the stellar parameters ([Fe/H], log $g$,  [$\alpha$/Fe] ) and the line-of-sight velocities are from the LSS-GAC DR4 catalog, and the proper motions and parallaxes  are from TGAS catalog \citep{Gaia16a, Gaia16b}.  We adopt the distance estimated by \cite{Astra16}, who applied a Bayesian model to derive the distance from the parallax, taking into account the Milky Way prior and systematic uncertainties in the Gaia catalog.    

\par Our initial sample was obtained by cross-matching
between the LAMOST and TGAS catalogs based on stellar position.  
Stars with large observational uncertainties were excluded from the sample. 
To ensure a sizable halo sample, we chose to use generous cuts  rather than stringent cuts on observational uncertainties.
There are in total more than 230,000 stars in common with SNR $\geq20$ and radial velocity uncertainties smaller than 10 kms$^{-1}$.  
Although it is not a very large sample, it can lend insights into the stellar kinematics in the solar neighborhood. 

\par For the following analysis, we transform the Galactic ($l$,  $b$) and distances for the stars into a Cartesian coordinate system ($X$, $Y$, $Z$).  We use a right-handed, Cartesian Galactocentric coordinate system  defined by the
following set of coordinate transformations:
\begin{align}
   & X =  R_\odot - D\,\cos(l)\,\cos(b) \nonumber \\
   & Y =  -D\,\sin(l)\,\cos(b) \\
   & Z =  D\,\sin(b), \nonumber
\end{align}
\noindent 
where $R_{\odot}=8.2$ kpc is the distance from the Sun
to the Galactic center \citep{Bland16},  $D$ is distance from the star to the Sun, and 
$l$ and $b$ are the Galactic longitude and latitude. Note that the $X$ axis is oriented toward $l=0^\circ$, the
$Y$ axis is oriented toward $l=90^\circ$ (the Sun's motion in the disk is toward
$l\sim90^\circ$), and the $Z$ axis toward the north Galactic pole. 

The tangential velocity $v$, is obtained from the proper motion
$\mu$ and the distance $D$ by
\begin{eqnarray}
 v  = 4.74\frac{\mu}{\rm mas\cdot yr^{-1}} \frac{D} {\rm kpc}~\rm km~s^{-1}.
\end{eqnarray}
The proper motions together with line of sight velocities are used to calculate the Galactic velocity components ($V_{X}=U$, $V_{Y}=V$, $V_{Z}=W$) and their errors, according to the formulae and matrix equations presented in \cite{Johnson87}. Here, we adopt a Local Standard of Rest velocity  $V_{\textrm{LSR}} = 220$ kms$^{-1}$, and the solar peculiar motion ($V_X^{\odot,{\rm pec}}, V_Y^{\odot,{\rm pec}},
V_Z^{\odot,{\rm pec}}$) = ($10.0 \rm ~km s^{-1}, 11.0~km s^{-1},
7.0~km s^{-1})$ \citep{Tian15, Bland16} and use these values to obtain
the Galactocentric velocity components:
\begin{align}
 &V_X = V_X^{{\rm obs}} + V_X^{\odot,{\rm pec}} \nonumber \\
 &V_Y = V_Y^{{\rm obs}} + V_Y^{\odot,{\rm pec}} + V_{\textrm{LSR}}\\
 &V_Z = V_Z^{{\rm obs}} + V_Z^{\odot,{\rm pec}} \nonumber 
\end{align}
We can now use this 6D phase space information to study the kinematics of local stars in the Milky Way.

\begin{figure*}
\includegraphics[width=1.0\hsize]{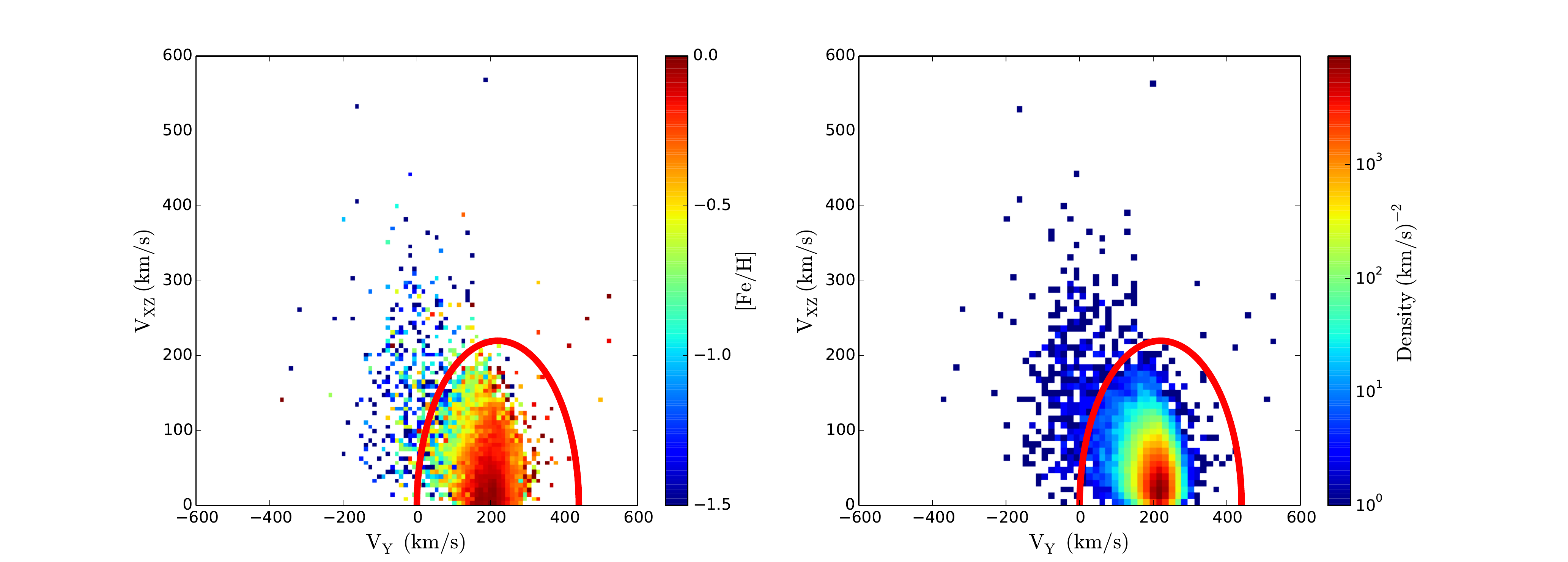}
\caption{Toomre diagram of stars in the solar neighborhood from the LAMOST and TGAS catalogs.  The dividing line between the components is marked with a red line. The left panel shows the distribution of sample stars with a measured metallicity in the Toomre diagram.  The color coding corresponds to the average metallicity of stars. Note that there are some halo stars that are quite metal-rich. The right panel shows the relative number of stars in each part of the diagram and the color coding corresponds to the number density of stars in each pixel.}
\label{figure1} 
\end{figure*}

\section{The local halo sample stars}

\par  The space distribution in the Toomre diagram has been widely used to  distinguish the thin-disk,
thick-disk, and halo stars \citep[e.g.,][]{Venn04, Bonaca17}.   Figure \ref{figure1} shows the Toomre diagram of stars in the solar neighborhood from the LAMOST and TGAS catalogs, where the $X$ axis represents the Galactocentric $Y$  velocity component, $V_{Y}$, whereas the $Y$ axis represents the perpendicular Toomre component, $\sqrt{V_{X}^2+V_{Z}^2}$.  As shown in Figure \ref{figure1},  
disk stars dominate a large overdensity at  $V_{Y} \sim 200$ kms$^{-1}$; the density of disk stars decreases smoothly in both directions from this $V_{Y}$ value, and does not populate retrograde orbits ($V_{Y} < 0$).  The halo stars on average have  $V_{Y} \sim 0$ kms$^{-1}$, as can be seen in the top portions of Figure \ref{figure1}.  
Following \cite{Nissen10} and \cite{Bonaca17}, we kinematically divide the sample stars into disk and halo components according to this Toomre diagram.  The halo stars are defined as having $|V-V_{\textrm{LSR}}| > 220 $ kms$^{-1}$, where $V_{\textrm{LSR}}= (0, 220, 0)$ kms$^{-1}$ in the Galactocentric Cartesian coordinates.  Here, we employ the halo definition following \cite{Bonaca17} , which is more conservative than similar cuts adopted by  \cite{Nissen10}.  The velocity cut ensures that the contamination from thick disk stars is minimized.  
The dividing line between the components is marked with a red line in Figure \ref{figure1}.   The left panel of Figure \ref{figure1} also shows the distribution of sample stars with a measured metallicity in the Toomre diagram, with color coding corresponding to the average metallicity of stars. The right panel shows the relative density of stars in each portion of the diagram and the color coding corresponds to the number density of stars in each pixel. In total, we identified 436 local halo stars within 3 kpc of the Sun. 

\par Surprisingly, there are many stars with disk-like metallicities ([Fe/H]$>-1.0$) found in the halo region of the Toomre diagram. Some  metal-rich stars are very far from the region of the diagram populated by disk stars; some are on strongly retrograde orbits, and some of those have large $V_{XZ}$ velocities as well.  
\cite{Bonaca17} found a similar result in their study using the Gaia data combined with RAVE and APOGEE spectroscopic surveys.  
The existence of metal-rich stars in kinematically-defined samples of halo stars implies that metallicity alone cannot be used to separate halo stars and disk stars.  

\par Since the metal-rich ([Fe/H]$>-1.0$) halo identified in the study has metallicities consistent with the thick disk, we therefore quantify the thick disk contamination to our halo sample under the assumption that 
the Galactic space velocities ($U$, $V$, and $W$) of the stellar populations in the thin disk, the thick disk, and the halo have Gaussian distributions:
\begin{eqnarray}
f(U,V,W)=k\cdot \textrm{exp}(\frac{U^2}{2\sigma_{U}^2}-\frac{(V-V_{\textrm{asym}})^2}{2\sigma_{V}^2}-\frac{W^2}{2\sigma_{W}^2}),
\end{eqnarray}
where 
\begin{eqnarray}
k=\frac{1}{(2\pi)^{3/2}\sigma_{U}\sigma_{V}\sigma_{W}}.
\end{eqnarray}
Here, $\sigma_{U}$, $\sigma_{V}$, and $\sigma_{W}$ are the characteristic velocity dispersions, and $V_{\textrm{asym}}$ is the asymmetric drift. 
The values of the three populations are listed in Table 1 \citep{Bensby03}.
\begin{table}[hbp]
\centering
\caption{ \upshape {Observed fraction of stars for the populations in the solar neighborhood, 
characteristic velocity dispersions ($\sigma_{U}$, $\sigma_{V}$, and $\sigma_{W}$) and the asymmetric drift ($V_{\textrm{asym}}$ )} }
\label{Table 1}
\begin{tabular}{cccccc}
\hline
\hline 
&X&$\sigma_{U}$& $\sigma_{V}$&$\sigma_{W}$ &$V_{\textrm{asym}}$\\
&  &  &  & [km/s] \\
\hline
Thin disk (D) & 0.94 & 35 & 20& 16 &-15\\
Thick disk(TD)& 0.06& 67& 38& 35 &-46\\
Halo (H) & 0.0015& 160& 90& 90& -220\\
\hline

\end{tabular}
\end{table}

To determine the probability that a given star belongs to a specific population, we multiply the probabilities from Eq. (4) by the observed fractions ($X$) of each population in the solar neighborhood.  We then obtain the relative probabilities for thick-disk-to-halo (TD/H) as follows: 

\begin{eqnarray}
\textrm{TD/H} = \frac{X_{\textrm{TD}}\cdot f_{\textrm{TD}}}{X_{H}\cdot f_{\textrm{H}}}
\end{eqnarray}

\par According to the thick disk and halo probability distributions, calculated with Eq. (4) and Eq. (5), only 35 stars with TD/H$>0.1$ are expected among the 436 stars in the halo sample.  Among all, only 23 stars have metallicity [Fe/H]$>-1.0$.   But in all halo sample stars, there has about 160 stars with  [Fe/H]$>-1.0$. 
So the thick disk still doesn't explain all metal-rich stars identified in this sample,  and particularly those with high velocities ($|V_{Y}|>260$) or retrograde orbits with high $V_{XZ}$.  This suggests that there exits a metal-rich halo component in addition to metal-poor inner and outer halo components.

\par
\begin{figure}
\includegraphics[width=1.0\hsize]{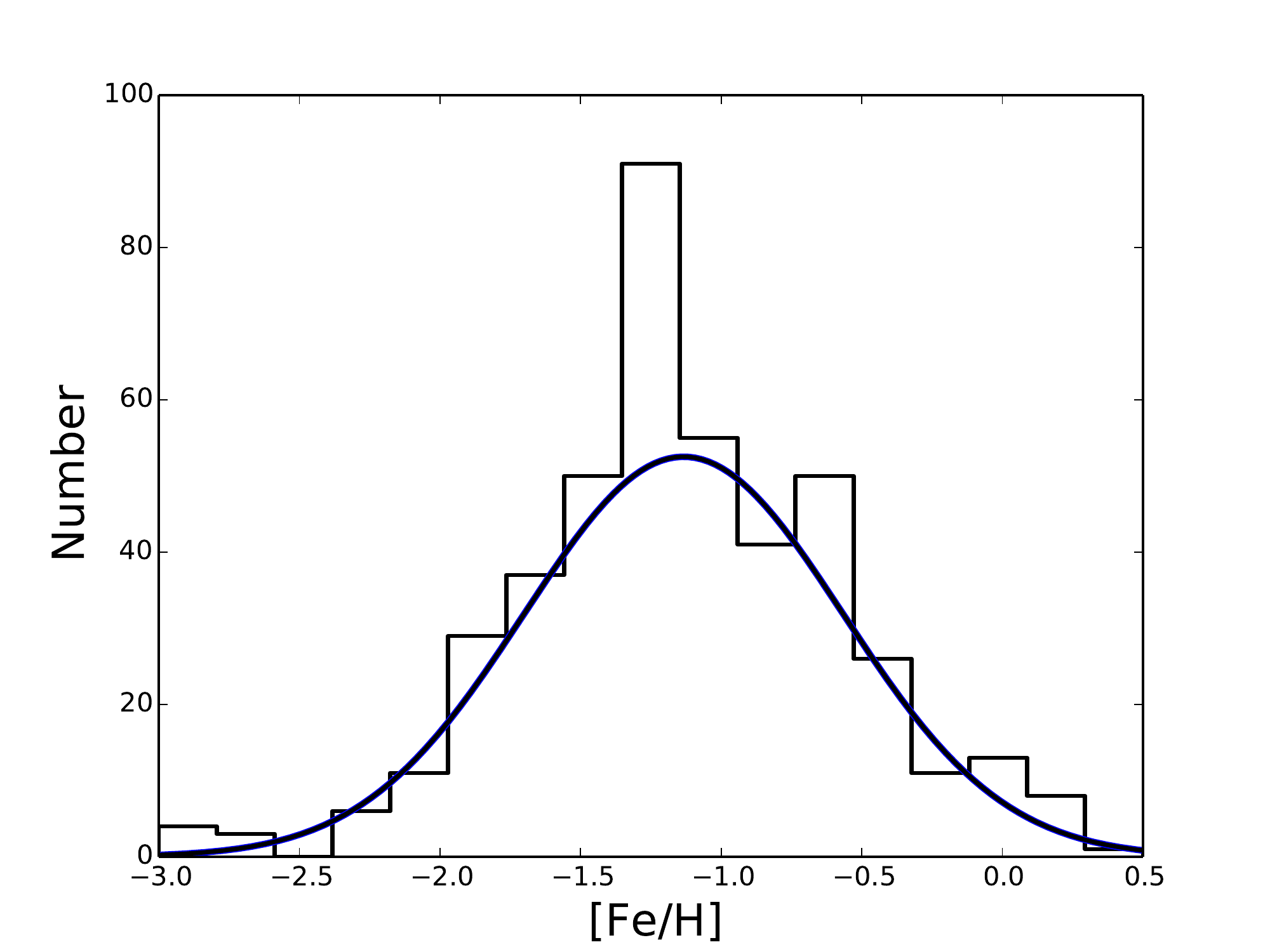}
\caption{The metallicity distribution of local halo stars is fitted by a Gaussian model with a peak near [Fe/H]$\sim-1.2$ and an tail extends out to super-solar values.  }
\label{figure2}
\end{figure}

 Figure \ref{figure2} presents the metallicity distribution of local halo stars; there is a wide metallicity distribution ranging [Fe/H]$\sim-3.0$ to [Fe/H]$\sim0.5$,
we fit the distribution with Gaussian model which peak at near [Fe/H]$\sim-1.2$ and an tail extends out to super-solar values.  \citet{Bonaca17} show the apparent bimodality in the metallicity distribution of RAVE-on halo stars is slightly more metal-poor, [Fe/H]$\sim-1.1$, than in the APOGEE sample, [Fe/H]$\sim-0.8$
in their Figure 2 , and approximately half of their halo sample is comprised of stars with [Fe/H]$>-1.0$. In this study, we found that about 30 percent of halo sample stars are metal-rich with [Fe/H]$>-1.0$ ,  but it is possible that the sample selection affect the metallicity distribution.

We decide the optimal number of Gaussian functions using the Bayesian information criterion (BIC):
\begin{align}
BIC= -2ln[L^{0}(M)] + klnN  \nonumber
\end{align}
where $L^{0}(M)$ represents the maximum value of the likelihood function of the model, $N$ is the number of data points, and $k$ represents the number of free parameters. More details about BIC can be found in \cite{Ivezic14}. As shown in Figure \ref{figure2}, we adopt one-peak Gaussian models to fit the metallicity distribution of local halo stars as the model with the lowest BIC.

\begin{figure*}
\centering
\includegraphics[width=0.8\hsize]{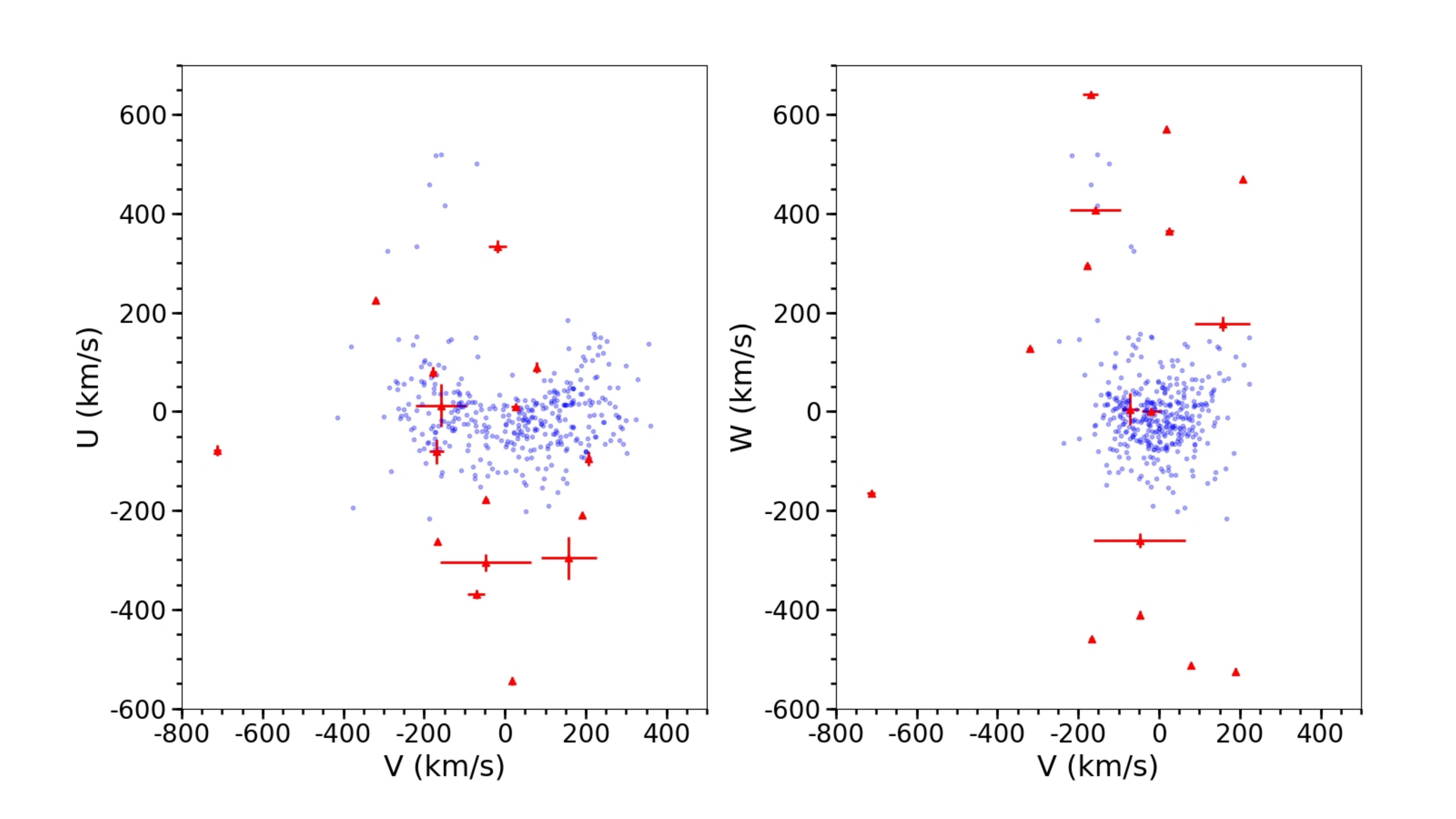} 
\caption{Velocity distribution of local halo stars in the LAMOST and TGAS catalog. 
The blue dots represent the halo sample stars selected from the Toomre diagram, and the red triangles represent the HiVel stars. }
\label{figure3}
\end{figure*}

\section {High Velocity Stars in the Local Stellar Halo}

\subsection{Selection of HiVel star candidates}

\par Before selecting HiVel star candidates, we removed stars with a higher likelihood of erroneous parameters. First, we selected only stars with calibrated T$_{\textrm{eff}}$  between  3500 and 8000 K and estimated log $g$ larger than 0.5 dex.  In addition, stars with extremely low metallicities ([Fe/H]$<-4.0$ dex) were discarded.
We then selected the HiVel stars with absolute Galactic radial velocity distribution greater than 200 kms$^{-1}$ in the final sample of halo stars.   
In order to derive reliable space velocities, we constrain the sample to stars with relative errors in the proper motions and distance smaller than 50 percent.  We subsequently derive the velocity in the Galactic rest frame V$_{\textrm{gsr}}$.   Our final selection criterion of 
V$_{\textrm{gsr}}>300$ km s$^{-1}$ gave us a HiVel candidate sample containing 16 stars.  

Atmospheric parameters and position for the HiVel stars can be found in Table 2.  Table 3 presents the space positions and velocities of the 16 HiVel stars. 
From the spatial distribution in Galactic coordinates, these HiVel stars are located in different Galactic directions. Therefore, it is possible that these HiVel stars have different origins.   Figure \ref{figure3} gives the space velocity distribution of our HiVel stars, showing that these local stars are not clumped in velocity.  

\begin{table*}[hbp]
\centering
\caption{ {\upshape Atmospheric parameters and positions for the 16 HiVel stars. }}
\label{Table 2}
\begin{tabular}{ccccccccccc}
\hline
\hline
Notation &            source-id &       l &      b & $\mu_{\alpha}$ $cos(\delta)$ &      $\mu_{\delta}$ &   RV$_{\odot}$ &  T$_{\mathrm{eff}}$ &  log(g) &            [Fe/H] &      [$\alpha$/Fe] \\
		 &					   &(deg)	&(deg)		　&(mas yr$^{-1}$)			 &(mas yr$^{-1}$)	  &(km s$^{-1}$)	&(K)		&			&		　＆		\\
\hline
  HiVel1 &  3266449244243890176 &  179.06 & -47.69 &            40.15 $\pm$ 1.52 &   -31.02 $\pm$ 0.84 &    407 $\pm$ 3 &                4900 &    2.28 &  -0.93 $\pm$ 0.12 &    0.2 $\pm$ 0.03 \\
  HiVel2 &  3893087103034072448 &  272.19 &  62.75 &            -6.86 $\pm$ 2.15 &    -6.01 $\pm$ 0.95 &    413 $\pm$ 4 &                4639 &    2.05 &  -1.58 $\pm$ 0.13 &   0.27 $\pm$ 0.03 \\
  HiVel3 &  1038229694366899200 &  157.03 &  42.34 &           -17.07 $\pm$ 0.48 &   -28.37 $\pm$ 0.82 &    301 $\pm$ 5 &                5014 &    2.71 &  -1.69 $\pm$ 0.12 &    0.2 $\pm$ 0.03 \\
  HiVel4 &  1544452200779441664 &  129.87 &  66.53 &           -14.11 $\pm$ 0.32 &      14.1 $\pm$ 0.4 &   -543 $\pm$ 5 &                5096 &    1.93 &  -2.41 $\pm$ 0.23 &   0.15 $\pm$ 0.06 \\
  HiVel5 &  4441393313920391936 &   27.11 &  28.04 &           -155.15 $\pm$ 1.1 &     8.46 $\pm$ 1.17 &   -420 $\pm$ 6 &                6477 &    4.12 &  -1.16 $\pm$ 0.12 &   0.28 $\pm$ 0.11 \\
  HiVel6 &  3662741856556426496 &  328.18 &  60.43 &           -229.0 $\pm$ 0.15 &    -80.12 $\pm$ 0.1 &    451 $\pm$ 6 &                6199 &    4.33 &  -1.86 $\pm$ 0.18 &                 - \\
  HiVel7 &  3962215976052842752 &  211.20 &  87.94 &           -49.38 $\pm$ 0.85 &    20.44 $\pm$ 0.74 &    466 $\pm$ 6 &                4254 &    1.83 &  -0.32 $\pm$ 0.15 &   0.13 $\pm$ 0.05 \\
  HiVel8 &   394095719362198784 &  118.24 & -13.80 &            70.48 $\pm$ 1.11 &   -14.18 $\pm$ 0.52 &   -599 $\pm$ 6 &                5158 &    2.50 &  -2.99 $\pm$ 0.23 &   0.13 $\pm$ 0.06 \\
  HiVel9 &  1245838311692516608 &   11.84 &  70.18 &           -66.73 $\pm$ 0.85 &    52.32 $\pm$ 0.48 &   -576 $\pm$ 7 &                5094 &    2.24 &  -2.85 $\pm$ 0.26 &   0.18 $\pm$ 0.08 \\
 HiVel10 &  1324910411958456064 &   53.19 &  42.22 &            12.45 $\pm$ 0.81 &    -16.4 $\pm$ 1.15 &   -670 $\pm$ 8 &                5671 &    3.27 &  -2.68 $\pm$ 0.22 &  -0.34 $\pm$ 0.18 \\
 HiVel11 &  2838296564529644288 &   97.32 & -35.76 &            -4.69 $\pm$ 1.96 &    -5.98 $\pm$ 0.46 &   -504 $\pm$ 9 &                5616 &    3.09 &  -1.57 $\pm$ 0.16 &   0.43 $\pm$ 0.14 \\
 HiVel12 &  1387977505574776320 &   62.11 &  56.16 &            -68.27 $\pm$ 0.5 &     41.3 $\pm$ 0.78 &  -533 $\pm$ 10 &                6473 &    3.96 &  -2.18 $\pm$ 0.28 &                 - \\
 HiVel13 &   866863316755386368 &  194.93 &  18.94 &           162.36 $\pm$ 0.16 &  -233.33 $\pm$ 0.11 &  -237 $\pm$ 10 &                6259 &    4.14 &   -1.9 $\pm$ 0.16 &   0.36 $\pm$ 0.12 \\
 HiVel14 &  3817216883707348352 &  249.93 &  58.72 &            14.31 $\pm$ 2.82 &   -31.89 $\pm$ 1.38 &    753 $\pm$ 3 &                3680 &    1.83 &   -0.32 $\pm$ 0.1 &   0.13 $\pm$ 0.03 \\
 HiVel15 &   645807259905057664 &  203.27 &  45.64 &            -23.12 $\pm$ 2.0 &     10.5 $\pm$ 1.37 &    818 $\pm$ 3 &                3694 &    1.83 &   -0.32 $\pm$ 0.1 &   0.13 $\pm$ 0.03 \\
 HiVel16 &  2086860081541233408 &   82.37 &  10.25 &            -7.17 $\pm$ 1.98 &   -13.75 $\pm$ 0.98 &  -962 $\pm$ 10 &                5733 &    4.36 &  -0.35 $\pm$ 0.17 &                 - \\
\hline
\end{tabular}
\end{table*}

\begin{table*}
\centering
\caption{ \upshape Spatial positions and Velocities of 16 HiVel stars }
\begin{tabular}{ccccccccccc}
\hline
\hline
Notation &    x &    y &    z &              U &              V &              W &  V$_{\mathrm{gsr}}$ &         D$_{\odot}$ &              $e$ &   Z$_{\mathrm{max}}$ \\
		&(kpc) &(kpc) &(kpc) &(km $s^{-1}$)  &(km $s^{-1}$)   &(km $s^{-1}$)  &(km $s^{-1}$)  &(pc) &     & (kpc)				\\
\hline
  HiVel1 &  9.0 & -0.0 & -0.9 &  -305 $\pm$ 18 &  -48 $\pm$ 113 &  -260 $\pm$ 14 &                 398 &  1199.5 $\pm$ 478.7 &  0.79 $\pm$ 0.03 &     45.3 $\pm$ 15.3 \\
  HiVel2 &  8.2 &  0.2 &  0.4 &     11 $\pm$ 6 &     26 $\pm$ 9 &    366 $\pm$ 6 &                 313 &   454.3 $\pm$ 244.9 &  0.54 $\pm$ 0.02 &      26.9 $\pm$ 1.3 \\
  HiVel3 &  9.0 & -0.4 &  0.8 &  -296 $\pm$ 43 &   157 $\pm$ 69 &   177 $\pm$ 15 &                 366 &  1224.1 $\pm$ 523.7 &  0.77 $\pm$ 0.04 &      28.4 $\pm$ 9.1 \\
  HiVel4 &  8.4 & -0.2 &  0.6 &    89 $\pm$ 11 &     78 $\pm$ 3 &   -512 $\pm$ 6 &                 469 &   687.2 $\pm$ 123.8 &  0.92 $\pm$ 0.01 &    195.4 $\pm$ 19.3 \\
  HiVel5 &  7.9 & -0.1 &  0.2 &   -368 $\pm$ 9 &   -71 $\pm$ 22 &     5 $\pm$ 32 &                 316 &    328.3 $\pm$ 53.7 &  0.93 $\pm$ 0.01 &       1.0 $\pm$ 2.9 \\
  HiVel6 &  8.1 &  0.1 &  0.2 &    13 $\pm$ 43 &  -158 $\pm$ 63 &    408 $\pm$ 5 &                 401 &    286.5 $\pm$ 66.4 &  0.74 $\pm$ 0.05 &     49.7 $\pm$ 17.9 \\
  HiVel7 &  8.2 &  0.0 &  0.4 &   -95 $\pm$ 14 &    207 $\pm$ 3 &    470 $\pm$ 6 &                 468 &    363.3 $\pm$ 56.8 &  0.91 $\pm$ 0.01 &    149.1 $\pm$ 12.5 \\
  HiVel8 &  8.3 & -0.2 & -0.1 &    226 $\pm$ 4 &   -320 $\pm$ 6 &    128 $\pm$ 2 &                 408 &    216.6 $\pm$ 11.2 &  0.75 $\pm$ 0.01 &      16.3 $\pm$ 1.9 \\
  HiVel9 &  8.2 & -0.0 &  0.1 &   -208 $\pm$ 3 &    190 $\pm$ 1 &   -525 $\pm$ 7 &                 550 &      72.2 $\pm$ 1.9 &   0.99 $\pm$ 0.0 &  1032.6 $\pm$ 130.6 \\
 HiVel10 &  8.1 & -0.2 &  0.2 &   -263 $\pm$ 4 &   -167 $\pm$ 5 &   -459 $\pm$ 6 &                 524 &    296.7 $\pm$ 22.8 &  0.96 $\pm$ 0.01 &    175.6 $\pm$ 27.4 \\
 HiVel11 &  8.3 & -0.4 & -0.3 &    80 $\pm$ 11 &   -177 $\pm$ 7 &    294 $\pm$ 7 &                 323 &   531.3 $\pm$ 294.6 &  0.51 $\pm$ 0.03 &      20.4 $\pm$ 1.3 \\
 HiVel12 &  8.2 & -0.1 &  0.1 &   -178 $\pm$ 4 &    -47 $\pm$ 5 &   -411 $\pm$ 8 &                 410 &     150.4 $\pm$ 8.1 &   0.8 $\pm$ 0.02 &      62.5 $\pm$ 5.9 \\
 HiVel13 &  8.4 &  0.1 &  0.1 &   334 $\pm$ 12 &   -19 $\pm$ 23 &      0 $\pm$ 6 &                 300 &    247.5 $\pm$ 18.3 &  0.98 $\pm$ 0.02 &       0.2 $\pm$ 1.3 \\
 HiVel14 &  8.3 &  0.2 &  0.3 &   -81 $\pm$ 24 &  -169 $\pm$ 19 &    641 $\pm$ 6 &                 644 &   334.7 $\pm$ 185.5 &                - &                   - \\
 HiVel15 &  8.4 &  0.1 &  0.2 &   -544 $\pm$ 8 &     18 $\pm$ 4 &    571 $\pm$ 6 &                 753 &    315.3 $\pm$ 85.8 &                - &                   - \\
 HiVel16 &  8.1 & -0.5 &  0.1 &   -78 $\pm$ 12 &  -712 $\pm$ 10 &   -166 $\pm$ 5 &                 734 &   513.1 $\pm$ 151.3 &                - &                   - \\
\hline
\end{tabular}
\end{table*}

\subsection{Chemical abundances of HiVel stars}
\par As discussed in detail by \citet{Gilmore98}, chemical abundances have been used to discern different components of the Galaxy.  
 Many recent surveys have shown that the different components of the Galaxy can be partially separated in [$\alpha$/Fe] vs. [Fe/H] distribution \citep{Nissen97, Stephens02, Nissen10, Lee11, Feltzing13, Haywood13}.  The distribution in [$\alpha$/Fe] space also provides information about the star formation rate in the stellar population. The  high [$\alpha$/Fe] found in halo and thick disk stars suggests that they formed in regions with a high star formation rate, so that only type II SNe contributed to their chemical enrichment. On the other hand, low$-\alpha$ stars originate in regions with relatively slow chemical evolution so that type Ia SNe have had time to form, and thus contribute iron to the interstellar medium before [Fe/H]$\sim-1.5$.  Since there is a higher iron abundance, the [$\alpha$/Fe] is lower at these higher metallicities \citep{Nissen10}.
Therefore, the abundance space of [$\alpha$/Fe] versus [Fe/H] is particularly useful in tracing the origin of individual stars \citep{Lee15}. 

\par
\begin{figure}
\includegraphics[width=1.0\hsize]{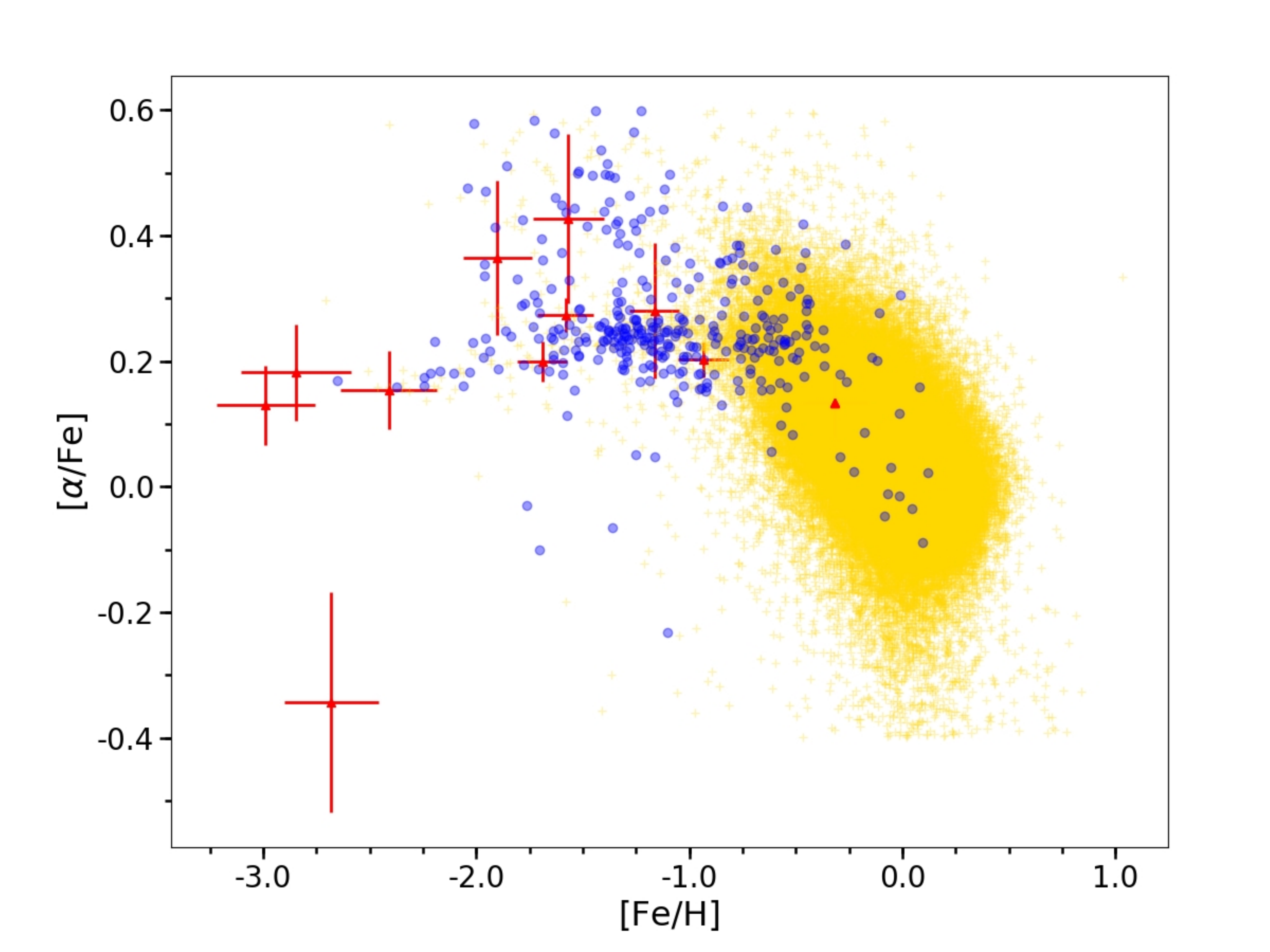} 
\caption{Chemical abundance distribution [$\alpha$/Fe] vs. [Fe/H] of halo stars in the TGAS-LAMOST sample. 
The red triangles represent the HiVel stars. The halo stars are shown individually as blue points and the disk stars shown as yellow plus signs for comparison.
The metal-poor halo is $\alpha$-enhanced, while the metal-rich halo follows the abundance pattern of the disk.}
\label{figure4}
\end{figure}

 Figure \ref{figure4} shows the chemical abundance distribution [$\alpha$/Fe] vs. [Fe/H] for all stars in this study. The red triangles represent the HiVel stars. The halo stars are shown individually as blue points and the disk stars are shown as yellow plus signs for comparison.  Notice that for the halo stars, there exist  high$-\alpha$ stars, with [$\alpha$/Fe] scatter from 0.2 to 0.6, and low$-\alpha$ stars, with [Fe/H]$>-0.5$ and a declining [$\alpha$/Fe] as a function of increasing metallicity.  
The metal-poor halo is $\alpha$-enhanced,   while the metal-rich halo follows the abundance pattern of the disk.
The large dispersion in the [$\alpha$/Fe] could result from the uncertainty of the individual [$\alpha$/Fe] estimates. The large uncertainty in the [$\alpha$/Fe] estimates, particularly for metal-poor stars, is a result of the relatively low resolution of LAMOST spectra.

\par We can see from Figure \ref{figure4} that our HiVel stars  are are metal-poor and $\alpha-$enhanced, except for HiVel7 ([Fe/H]$=-0.32$, [$\alpha$/Fe]=0.13).  HiVel7 is kinematically consistent with the halo but chemically consistent with disk.  \citet{Hawkins15} use RAVE data to discover one such metal-rich halo star and they consider it has likely been dynamically ejected into the halo from the Galactic thick disk.   \citet{Purcell10} also suggested that the inner parts of galactic stellar halos contain ancient disk stars, which could be ejected into the halo by the merging of satellite galaxies.  \citet{Zolotov09}  found that stars formed out of accreted gas in the inner 1 kpc of the Galaxy can be displaced into the halo through a succession of mergers.  In contrast, the high$-\alpha$ population might have formed as the first stars in a dissipative collapse of a proto-Galactic gas cloud \citep{Gilmore89, Schuster06}.

\begin{figure}
\includegraphics[width=1.0\hsize]{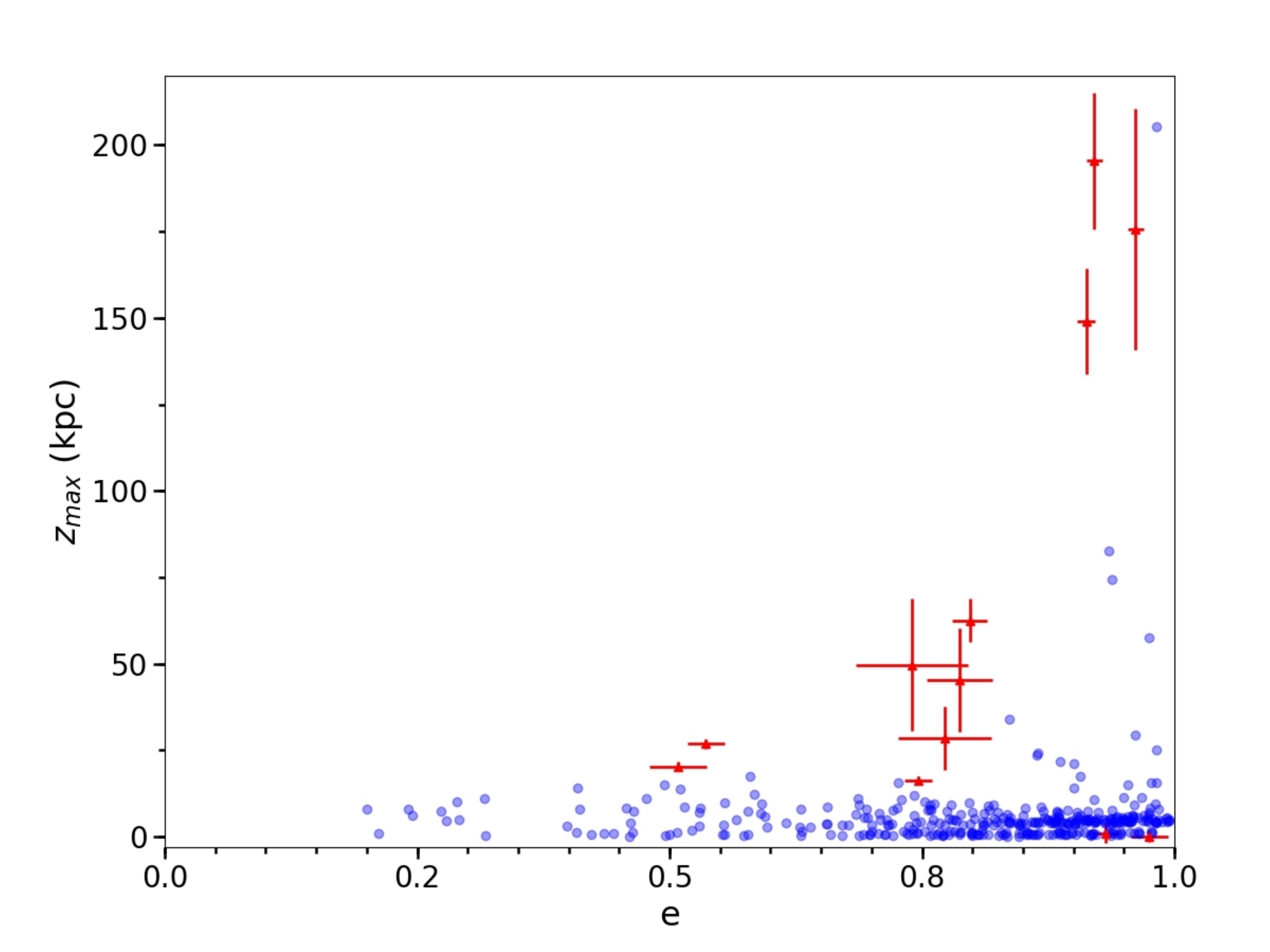} 
\caption{Eccentricity, $e$, as a function of the maximum height above the Galactic plane, Z$_{\textrm{max}}$. The red triangles represent the HiVel stars. The halo stars are shown individually as blue points for comparison. }
\label{figure5}
\end{figure}

\begin{figure*}
\includegraphics[width=1.0\hsize]{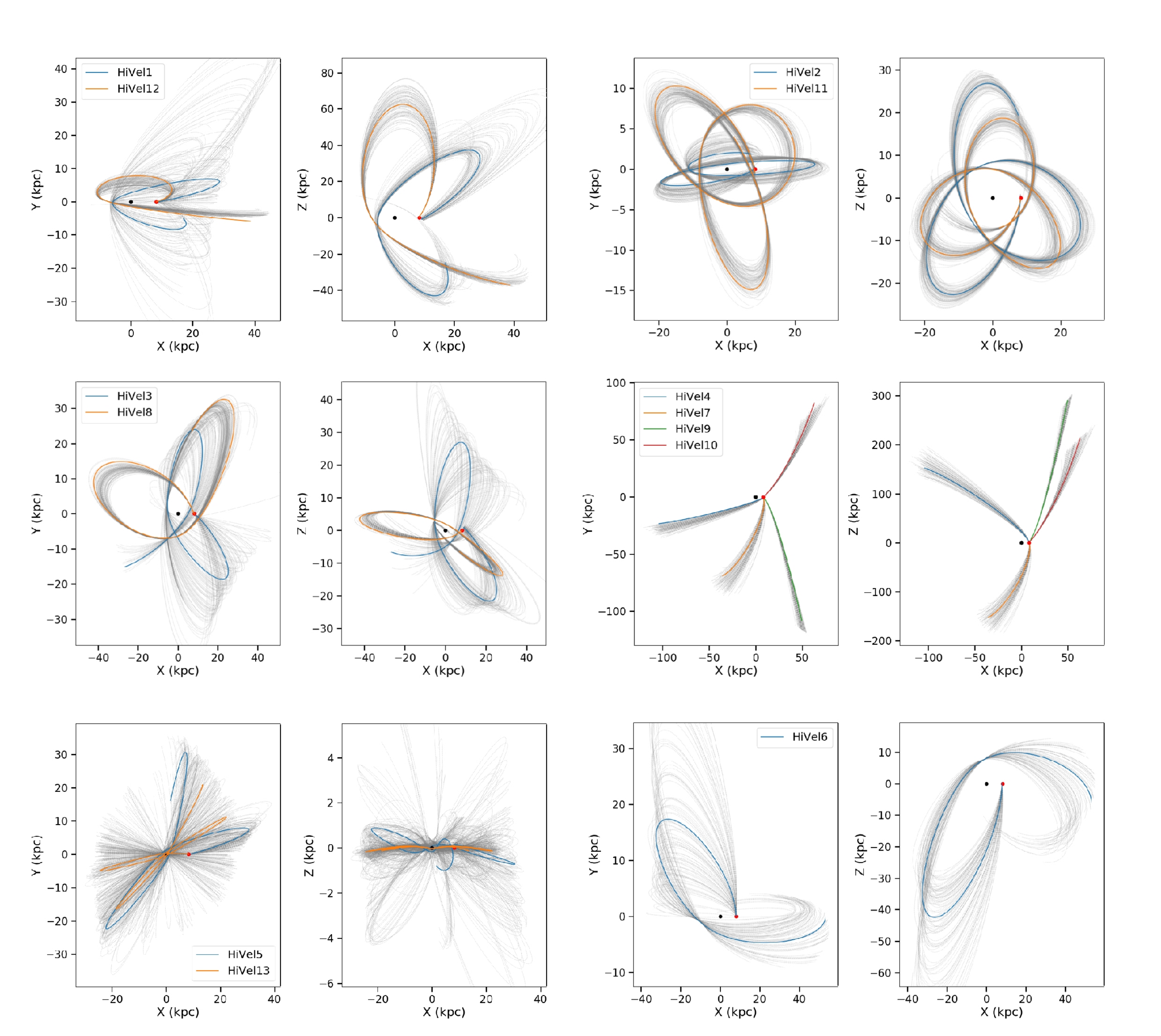} 
\caption{1 Gyr backwards orbit of the individual HiVel stars in $XYZ$ Galactocentric Coordinates. 
The red dot represents the current position and the black dot represents the Galactic Center. The thin grey lines show 100 orbits drawn at random from the uncertainties in the positions and velocities of each HiVel star, showing the uncertainty in the orbits.}
\label{figure6}
\end{figure*}

\subsection{Orbits of the HiVel stars}

\par For each of the stars in our local halo sample, we investigate their
orbital properties by adopting a Galaxy potential model. 
In this study, we use a recent Galactic potential model provided in \citet{McMillan17}. 
This new model includes components that represent the contribution of the cold gas discs near the Galactic plane, as well as thin and thick
stellar discs, a bulge component and a dark-matter halo.
We estimated the maximum distance above the Galactic plane (denoted Z$_{\textrm{max}}$) and the eccentricity, $e$, from the orbital integration.  
The eccentricity is defined as $e=(r_{\rm apo} - r_{\rm peri})/(r_{\rm
apo} + r_{\rm peri})$, where $r_{\rm peri}$ denotes the closest
approach of an orbit to the Galactic center (i.e., the perigalactic
distance), and ${r_{\rm apo}}$ denotes the farthest extent of an
orbit from the Galactic center. 
Figure \ref{figure5} shows the $e$ - Z$_{\textrm{max}}$ plane, which allows us to characterize the orbits of our sample stars; $e$ describes the shape of the orbit and Z$_{\textrm{max}}$ describes the amplitude of the vertical oscillations \citep{Boeche13}.  
Figure~\ref{figure5} shows that the HiVel stars have $e>0.5$ and most have Z$_{\textrm{max}}>10$ kpc, reaffirming that these stars are decidedly not associated with a disk \citep{Schuster88, Ryan03, Schuster06}.  Their orbits would take them into the outer halo.  However, there are two stars with Z$_{\textrm{max}}<2$ kpc and eccentricities $e>0.6$.

\par Next, we considered the origin of these HiVel stars.
As outlined in the introduction, the high velocities of HiVel stars indicate that they may have originated from a strong dynamical interaction with a BH (or BHs) in the GC or from a nearby galaxy \citep{Sherwin08}.

\par To understand the origin of high velocity stars, we calculated the backwards orbits of individual stars  to see if they converge somewhere, and in particular whether they originate from the Galactic center.
We did this by integrating the orbit in our Galactic potential model, starting with the current position of each star and the negative of its current velocity.  
In all 16 HiVel stars,  3 HiVel stars (Hivel14, HiVel15 and HiVel16) are unbound due to their very high velocity and only the orbit of 13 HiVel stars could be determined. 
Figure \ref{figure6} gives the derived backward orbits for 13 HiVel star, integrated back 1 Gyr.  The red dot represents the present position, and the black dot represents the Galactic Center. As seen in Figure \ref{figure6}, a few high velocity stars appear to originate from the Galactic center, but others are not consistent with a GC origin, and must be produced by another mechanism.

\par According to the orbital integration of HiVel stars ( shown in Figure \ref{figure6}), HiVel4, HiVel7, HiVel9 and HiVel10 could not originate in the Galactic Center. Combining their chemical and orbit information, we conclude that HiVel4, HiVel9 and HiVel10 could originate from the tidal debris of an accreted and disrupted dwarf galaxy \citep{Abadi09} or globular cluster. While for HiVel7, the disrupted dwarf galaxy or globular cluster explanation are unlikely due to  the chemical composition of the stars.  The star likely originates in the thick disk where one would expect a richer metallicity and lower $\alpha$-abundance.   HiVel5 ,HiVel13, HiVel3 and HiVel8 possiblely are ejected from near the Galactic Center.   For the rest of the HiVel stars (HiVel1, HiVel2, HiVel11 and HiVel12), it is possible that they were kicked from the Galactic disk. The mechanism by which these stars were ejected from the disk, namely binary supernova explosion, interaction of a dwarf galaxy or a globular cluster with the disk, or interaction between multiple stars or other gravitational mechanisms, is unclear.

\section{Conclusions and summary}

Based on the first year of Gaia data combined with observations from ground-based spectroscopic survey LAMOST DR4,  we analyzed  a sample of local halo stars within $\sim3$ kpc  of the Sun.   For the kinematically identified local halo stars, we found  $30 \%$ of them have [Fe/H] $>-1.0$, which is more metal-rich than expected in the inner halo.   For the halo stars, there also exist  high$-\alpha$ stars, with [$\alpha$/Fe] scatter from 0.2 to 0.6, and low$-\alpha$ stars, with [Fe/H]$>-0.5$ and a declining [$\alpha$/Fe] as a function of increasing metallicity. 
For each of the stars in our local halo sample, we also adopt an Galactic potential model to derive their orbital parameters, particularly Z$_{max}$ and eccentricity, to study the kinematics.

From this halo sample, 16 high velocity stars are identified.
We studied the metallicity and [$\alpha$/Fe] distribution of our HiVel stars.  While most of the HiVel stars are metal-poor, there are several stars that have 
metallicity above $-0.5$ dex.  These stars, while having kinematics that resemble halo stars, have disk-like metallicity and thus don't conform to the the rest of the HiVel stars.
To understand the origin of high velocity stars, we calculated the backwards orbits of each HiVel stars and found that there is a few high velocity stars which appear to originate from the Galactic center, but several stars are not consistent with a GC origin and could have been kicked up from the Galactic disk.

\section*{Acknowledgements}

\par  We thank especially the referee for insightful comments and suggestions, which have improved the paper significantly. 
This work was supported by joint funding for Astronomy by the National Natural Science Foundation of China and the Chinese Academy of Science, under Grants U1231113.  This work was also by supported by the Special funds of cooperation between the Institute and the University of the Chinese Academy of Sciences, and China Scholarship Council (CSC).  HJN acknowledges funding from NSF grant AST 16-15688. Funding for SDSS-III has been provided by the Alfred P. Sloan Foundation, the Participating Institutions, the National Science Foundation, and the U.S. Department of Energy Office of Science. 
This project was developed in part at the 2016 NYC Gaia Sprint, hosted by the Center for Computational Astrophysics at the Simons Foundation in New York City. The Guoshoujing Telescope (the Large Sky Area Multi-Object Fiber Spectroscopic Telescope, LAMOST) is a National Major Scientific Project built by the Chinese Academy of Sciences. Funding for the project has been provided by the National Development and Reform Commission. LAMOST is operated and managed by the National Astronomical Observatories, Chinese Academy of Sciences. This work has made use of data from the European Space Agency (ESA) mission Gaia (http://www.cosmos.esa.int/gaia), processed by the Gaia Data Processing and Analysis Consortium (DPAC, http://www.cosmos.esa.int/web/gaia/dpac/consortium). Funding for DPAC has been provided by national institutions, in particular the institutions participating in the Gaia Multilateral Agreement.

\end{document}